\documentclass[conference]{IEEEtran}
\ifCLASSINFOpdf
\else
\fi
\usepackage{amsmath}
\usepackage{amssymb}
\usepackage{amsfonts}
\usepackage{psfrag}
\usepackage{epsfig} 
\usepackage{algorithm}
\usepackage{algorithmic}

\newcommand{\B}[1]{\boldsymbol{#1}}
\begin{document}
%
\title{Blind Estimation of Sparse Multi-User Massive MIMO Channels}

\author{\IEEEauthorblockN{Amine Mezghani and A. Lee Swindlehurst}
\IEEEauthorblockA{Center for Pervasive Communications and Computing (CPCC)\\
Department of EECS, University of California, Irvine\\
Irvine, CA 92697, USA\\
Email: \{amezghan, swindle\}@uci.edu} }


%


\maketitle

\begin{abstract}
We provide a maximum likelihood formulation for the blind estimation of massive mmWave MIMO channels while taking into account their underlying sparse structure. The main advantage of this approach is the fact that the overhead due to pilot sequences can be reduced dramatically especially when operating at low SNR per antenna. Thereby, the sparsity in the angular domain is exploited as a key property to enable the unambiguous blind separation between  user's channels. On the other hand, as only the sparsity is assumed, the proposed method is robust with respect to the statistical properties of the channel and data and allows the estimation in rapidly time-varying scenarios and eventually the separation of interfering users from adjacent base stations. Additionally, a performance limit is derived based on the clairvoyant Cram\'er Rao lower bound.  Simulation results demonstrate that this maximum likelihood formulation yields superior estimation accuracy with  reasonable computational complexity and limited model assumptions.

\end{abstract}


%
\IEEEpeerreviewmaketitle

\section{Introduction}
Channel estimation is recognized as one of the key problems for developing the fifth generation of communication systems \cite{eetimes}. In particular, estimating massive MIMO millimeter wave (mmWave) channels is challenging due to larger dimensions, larger bandwidths, hardware imperfections and faster temporal variations. In addition, such systems are expected to operate at low SNR values per antenna caused by several factors like increased path-loss, hardware restrictions of power amplifiers, larger noise bandwidths and smaller antenna sizes, which, together with the issue of pilot-contamination, renders common pilot based estimation methods inefficient and even impossible. \\

 Previous works have exploited the sparsity of mmWave channels in the angle and delay
domains to design pilot-based channel estimation schemes \cite{You_2016,Schniter_2014}. Other works have considered pilot based channel and/or channel subspace  estimation in the context of hybrid MIMO mmWave systems with analog preprocessing \cite{Haghighatshoar_2016} and in the context of quantized MIMO mmWave systems with one-bit receivers \cite{Mo_2014}.  A maximum likelihood approach for 
blind and semi-blind estimation of massive MIMO mmWave channel has been presented in \cite{Neumann_2015} for Rayleigh fading channel models. Joint Bayesian channel-and-data estimation has been considered and analyzed in \cite{Wen_2015,Wen_Wu_2015,Steiner_2016} where a large improvement has been obtained compared to training-based methods. However, this approach requires an iterative solution with significant algorithmic complexity and generally accurate assumptions of the channel and data prior distributions, and convergence and optimality still  cannot be guaranteed. To address this issue, we present a  maximum likelihood approach for blind mmWave channel estimation that, unlike \cite{Neumann_2015}, takes into account the sparsity of these channels. Under this key property, we show then that reliable estimation is possible at low SNR per antenna and unambiguous separation between users is still possible even though their channels are not orthogonal and channel and data prior distributions are not available. \\

\emph{Notation:}
Vectors and matrices are denoted by lower and upper case italic bold letters.  The operators $(\bullet)^\mathrm {T}$, $(\bullet)^\mathrm {H}$, $\textrm{tr}(\bullet)$ and $(\bullet)^*$ stand for transpose, Hermitian (conjugate transpose), trace, and complex conjugate, respectively.  $\B{1}_M$ and $\B{I}_M$ stand for the all ones vector and identity matrix of size $M$, respectively. $\boldsymbol{x}_i$ is the $i$-th column of a  matrix $\B{X}$ and $\left[\B{X}\right]_{i,j}$ denotes the ($i$th, $j$th) element, while $x_i$ is the $i$-th element of the vector $\B{x}$.  Finally, we represent the Hadamard (element-wise) and the Kronecker product of vectors and matrices by the operator "$\circ$" and "$\otimes$", respectively. 
\section{Channel Model}
In mmWave transmission, the wireless  propagation channel can be described by a sparse scattering  model, where  the $N$-dimensional channel vector $\B{h}_k$ of user $k$ to the base station consists of the superposition of only $L_k \ll N$ multi-path components, mostly including the line-of-sight (LOS) and some reflected paths. Assuming a uniform linear array (ULA) with spacing $d$, that is
\begin{equation}
\B{h}_k= \sum_{\ell=1}^{L_k} s_{\ell,k} \B{a}(\theta_{\ell,k}),
\end{equation}
where $s_{\ell,k}$ are the path coefficients (including path phase and strength),  $\B{a}(\theta_{\ell,k})$ the array response for the angle-of-arrival (AoA) $\theta$, i.e.,
\begin{equation}
\B{a}(\theta)= \left[1,\cdots, {\rm e}^{{\rm j}\frac{2 \pi}{\lambda}dn \sin(\theta)} ,\cdots,  {\rm e}^{{\rm j}\frac{2 \pi}{\lambda}d(N-1) \sin(\theta)} \right]^{\rm T}.
\end{equation}
Apart of sparsity, no further assumption is made about the channel's statistical properties for the derivation of the estimation method and the analysis.\footnote{Generalization of the approach to the frequency selective case is possible by way of taking into consideration the sparsity in the delay domain. For simplicity, however, we concentrate here on the flat fading case.} For the simulation part, we assume for simplicity that the AoAs $\theta_{\ell,k}$ are random and uniformly distributed between $0$ and $\pi$, while the multi-path coefficients $s_{\ell,k}$ are drawn from a complex Gaussian distribution with unit variance. 
\section{Blind Channel Estimation: Conventional Subspace Method}

We consider the following block fading channel model with $K$ single antenna users  and $N$  receive antennas in the uplink. During the coherence time (in symbols) $T$, the base station receives the following data block
\begin{equation}
    \B{Y} = \B{H} \cdot \B{X} + \B{N},
\end{equation}
where $\B{N} \in \mathbb{C}^{N \times T}$ is the noise matrix  having i.i.d. elements with unit variance, $\B{H} =[\B{h}_1,\ldots,\B{h}_K] \in \mathbb{C}^{N \times K}$ comprises the user channels  $\B{h}_k$, $k=1, \ldots,K$ assumed to be unknown and $\B{X} =[\B{x}_1, \ldots, \B{x}_T] \in \mathbb{C}^{K \times T}$ is the transmit data during the coherence time.  Assuming the data from the users $x_{k,t}$ are Gaussian distributed  with variance $\rho$, then the conditional distribution of the receive matrix $\B{Y}$ given $\B{H}$ can be expressed as a multivariate Gaussian distribution with covariance matrix   $(\rho \B{H}\B{H}^{\rm H} + \B{I}) \otimes \B{I}_T $
\begin{equation}
     p(\B{Y}|\B{H}) = \frac{{\rm exp}\left( - {\rm tr} \left( \B{Y}^{\rm H} \left( \rho \B{H}\B{H}^{\rm H} + \B{I} \right)^{-1} \B{Y} \right) \right)}{\pi^{N\cdot T}\left|\rho \B{H}\B{H}^{\rm H} + \B{I} \right|^T},
\end{equation}
where $\rho$ represents the SNR. \\

 One common solution for the blind estimation of $\B{H}$ that maximizes the conditional distribution is given by \cite{Neumann_2015}
\begin{equation}
    \hat{\B{H}} =  {\rm argmax}_{\B{H}} \quad p(\B{Y}|\B{H}) = \frac{1}{\sqrt{T\rho}}  \B{U}_{1:K}  \sqrt{[\B{\Sigma}_{1:K}- \B{I}]_+},
\label{sol_subspa}
\end{equation}
where $\B{U}_{1:K} $ are the $K$ eigenvectors corresponding to the $K$ largest eigenvalues $\B{\Sigma}_{1:K}$ of the matrix $\B{Y}\B{Y}^{\rm H}=\B{U}\B{\Sigma}\B{U}^{\rm H}$ and $[a]_+=\max(a,0)$. It should be noticed that this solution is not unique and that multiplication from the right with any unitary matrix will also provide another valid solution.  The particular channel estimate in (\ref{sol_subspa}) is characterized by the fact that the users are assumed to be orthogonal to each other. Therefore the quality of the subspace-based estimate strongly depends on this  assumption, which requires an extremely high number of antennas. In the following section, we provide a modification of the method exploiting the sparsity of the propagation scenario that can relax this assumption.  
\section{Blind Sparse Channel Estimation}

For the case of mmWave massive MIMO, the number of multi-path components from each user to the base station is usually much less than the number of antenna elements $N$. Therefore, assuming a uniform linear array (ULA), the channel can be represented as
\begin{equation}
\B{H}= \B{F} \cdot \B{S},
\end{equation}
where $\B{F} $  is the unitary Discrete Fourier Transform (DFT) matrix of size $N$ and $\B{S}$ is the sparse matrix representing the coefficients of the different multi-path components. Strictly speaking, the matrix $\B{S}$ is not perfectly sparse since the path directions from the users to the base stations do not correspond exactly to discrete  directions defined by the DFT, resulting in a clustered type of sparsity (known as the leakage phenomenon). Nevertheless, the assumption of sparsity becomes more valid the higher the number $N$ of base station antennas.

Based on these assumptions we can state the following regularized maximum likelihood problem for estimating the channel:
\begin{equation}
\begin{aligned}
    &\max\limits_{\B{S}} L(\B{S})- \lambda \left\| \B{S} \right\|_{1,1} = \\
   &\!\! -\! {\rm tr} \! \left(\! \! \B{Y}^{\rm H} \B{F} \! \left( \!\rho \B{S}\B{S}^{\rm H} \!+\! \B{I} \!\right)^{-1} \!\! \B{F}^{\rm H}  \B{Y} \!\! \right) \!-\!T \log\left|\rho \B{S}\B{S}^{\rm H}\! +\! \B{I} \right| \!-\! \lambda \! \left\| \B{S} \right\|_{1,1} \!,
    \label{opt_S}
\end{aligned}    
\end{equation}
where the $\ell_{1,1}$ matrix norm $\left\| \B{S} \right\|_{1,1} = \sum_{n,k} |s_{n,k}|$ is used to encourage sparse solutions with regularization parameter $\lambda$. The advantage of this formulation is that apart from the channel sparsity, no further assumption is made on the data and the channel's distributions, which provides robustness  and allows for the estimation of even the channels of interfering users from adjacent base stations. \\

Unfortunately, the problem in (\ref{opt_S}) is non-convex since the cost function is not concave. Nevertheless, it has been observed that in many cases solving a non-convex problem locally using efficient gradient based methods \cite{Li_2016} can be very successful for solving the problem in practice, even though mathematically not rigorous. Therefore, we use the gradient descent based iterative thresholding algorithm to determine a local optimal solution, as derived in the Appendix (see also \cite{Zymnis_2010,Hale_2008}):
\begin{equation}
\begin{aligned}
 &\B{S}_{\ell+1}= \\
 &{\rm exp} ( {\rm j} \angle (\B{S}_\ell -\mu \B{\Delta})) \circ \max \left( {\rm abs}( \B{S}_\ell -\mu \B{\Delta} ) - \mu \frac{\lambda}{2} \B{1} \cdot \B{1}^{\rm T}, \B{0} \right),
 \end{aligned}
 \label{fixed_iter}
\end{equation}
where the phase and the absolute value operations symbolized by $\angle(\bullet)$ and ${\rm abs}(\bullet)$, respectively, are applied element-wise to the matrix and the gradient is given by  
\begin{equation}
\begin{aligned}
   \B{\Delta}  =& - \frac{\partial L(\B{S})}{\partial \B{S}^*} \\
   =& -\rho \left( \rho \B{S}\B{S}^{\rm H} + \B{I} \right)^{-1} \B{F}^{\rm H}  \B{Y} \B{Y}^{\rm H} \B{F}  \left(\rho\B{S}\B{S}^{\rm H} + \B{I} \right)^{-1}  \B{S} +  \\
    & \quad\quad\quad\quad\quad\quad\quad\quad\quad\quad\quad\quad T\cdot \rho  \left(\rho\B{S}\B{S}^{\rm H} + \B{I} \right)^{-1}  \B{S}. 
\end{aligned}
\label{gradient}
\end{equation}
As initialization for the iterative algorithm we take the subspace solution (\ref{sol_subspa}). The iterative method is summarized in Algorithm~\ref{Blind_alg}. \\

 We note that the solution of the optimization problem (\ref{opt_S}) is not unique, in the sense that any transformation of the form
\begin{equation}
  \B{H}'=\B{H}\B{\Phi}\B{\Pi},
  \label{perm}
\end{equation}
with any diagonal phase shift matrix  $\Phi$ and any permutation matrix  $\B{\Pi}$ provides an equally valid solution, a fact that reflects the non-uniqueness of assigning the channels to the user's indices. These ambiguities in terms of phase shift and user assignment can be resolved easily by using the structure of the modulation scheme (e.g. QPSK) and information from the higher layers or by including a short training phase.
\begin{algorithm}[t]
\caption{Blind $\ell_1$ Regularized Channel Estimation}
\label{Blind_alg}
\begin{algorithmic}[1]
\STATE \textbf{Initialize:} $ \B{U}\B{\Sigma}\B{U}^{\rm H} \leftarrow \B{F}^{\rm H}\B{Y}\B{Y}^{\rm H}\B{F}$ \\
$\B{S}_0 =  \frac{1}{\sqrt{T\rho}}  \B{U}_{1:K}  \sqrt{[\B{\Sigma}_{1:K}- \B{I}]_+}$
\\     
$\mu > 0 $, $0<\beta <1$, $l\leftarrow 0$      
\REPEAT
\STATE  $\ell \leftarrow \ell+1$
\STATE Compute $\B{\Delta}_{\ell-1}$ from (\ref{gradient})
\STATE Gradient update:\\
 $\B{S}_\ell \leftarrow \B{S}_{\ell-1}-\mu \B{\Delta}_{\ell-1}$\\
\STATE Thresholding: \\
 $ \B{S}_{\ell}  \leftarrow  {\rm exp} ( {\rm j} \angle (\B{S}_\ell)) \circ \max \left( {\rm abs}( \B{S}_\ell  ) - \mu \frac{\lambda}{2} \B{1} \cdot \B{1}^{\rm T}, \B{0} \right)$
\IF{$L(\B{S}_{\ell-1})- \lambda \left\| \B{S}_{\ell-1} \right\|_{1,1}  > L(\B{S}_{\ell})- \lambda \left\| \B{S}_{\ell} \right\|_{1,1} $} 
\STATE $\mu \leftarrow \beta\mu,\quad \ell\leftarrow\ell-1$\ENDIF
\UNTIL{desired accuracy for $\B{S}$ is achieved}
\STATE DFT conversion:\\
   $\hat{\B{H}}=\B{F}\cdot \B{S}$
\end{algorithmic}
\end{algorithm}
\section{Clairvoyant Cram\'er Rao lower bound (CRLB)}
For given channel, i.e., $\B{H}=[\cdots \B{h}_k\cdots]= [\cdots \B{F} \B{s}_k\cdots]$, and assuming that the right singular vectors are perfectly known - to ensure the uniqueness of the maximum likelihood solution - then the Fisher information matrix, leading to the so-called clairvoyant Cram\'er Rao lower bound (with ``genie" side information), can be written for a multivariate Gaussian distribution with zero mean as \cite{Kay_1993}

\begin{equation}
\begin{aligned}
 &\!\!\!\!\!\!\!\!\left[\B{J}\right]_{N(k-1)+i,N(k'-1)+i'}  \\
 &=T \cdot {\rm tr} \left(\B{Q}^{-1} \frac{\partial \B{Q}}{\partial s_{k,i}^*} \B{Q}^{-1} \frac{\partial \B{Q}} { \partial s_{k',i'} } \right) \\
 &= T \cdot \rho^2 \cdot {\rm tr} \left(\B{Q}^{-1} \B{F} \B{s}_k {\bf e}_i^{\rm T} \B{F}^{\rm H} \B{Q}^{-1} \B{F} {\bf e}_{i'} \B{s}_{k'}^{\rm H} \B{F}^{\rm H}  \right)  \\
 &=T \cdot \rho^2\cdot {\bf e}_i^{\rm T} \B{F}^{\rm H} \B{Q}^{-1} \B{F} {\bf e}_{i'}     \cdot \B{s}_{k'}^{\rm H} \B{F}^{\rm H} \B{Q}^{-1} \B{F} \B{s}_k,
\end{aligned}
\end{equation}
 with $\B{Q}=\rho \B{H}\B{H}^{\rm H}+\B{I}$. This can be written in a compact way as
 \begin{equation}
  \B{J}= T \cdot \rho^2 \cdot \B{H}^{\rm H} \B{Q}^{-1} \B{H} \otimes \B{F}^{\rm H} \B{Q}^{-1} \B{F} .
     \label{fisher}
 \end{equation}
 At low SNR, i.e.,  $\B{Q}\approx \B{I}$, this can be approximated as 
 \begin{equation}
  \B{J}\stackrel{\rho \ll 1}{\approx} T \cdot \rho^2 \cdot \B{H}^{\rm H}  \B{H} \otimes \B{I}_N.
 \end{equation}
 
Additionally, knowing the sparsity structure (support) of the channel, i.e.,  the indices of the non-zero elements of $\B{s}_k$, $\mathcal{S}=\{N(k-1)+i| s_{k,i} \neq 0 \}$, the reduced Fisher information matrix can be obtained by taking the rows and the columns given by the subset $\mathcal{S}$ as follows
 \begin{equation}
   \tilde{\B{J}}=\B{J}_{\mathcal{S},\mathcal{S}}.
   \label{fisher2}
  \end{equation}
\section{Simulation results}
\begin{figure}[h]
\psfrag{x}[c][c]{$\eta$}
\psfrag{y}[c][c]{${\rm Pr}\left(\frac{|\hat{\B{h}}_k^{\rm H} \B{h}_k |}{\left\|\B{h}_k  \right\|_2\left\|\hat{\B{h}}_k  \right\|_2} \geq \eta \right)$}
\centerline{\epsfig{figure=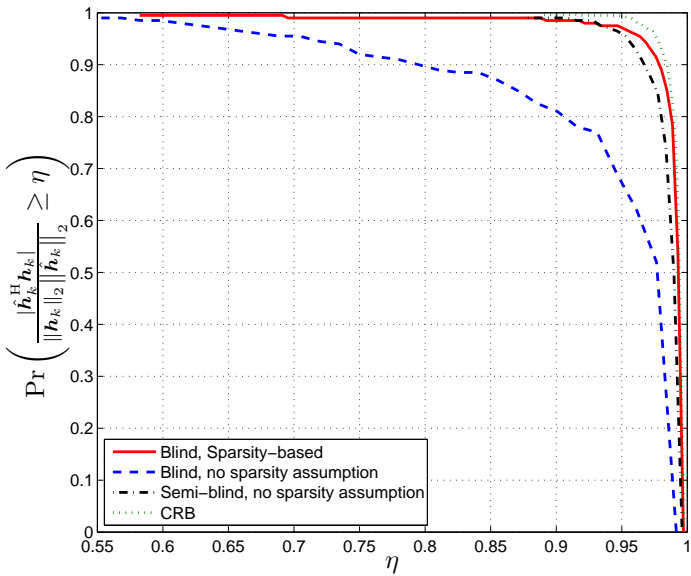,width=9cm}}
\caption{Estimation Performance for $\rho=-12$dB, $N=32$, $K=2$, $L=3$, $T=1000$.}
\label{fig_est}
\end{figure}

As a performance measure for  evaluating the performance of the presented algorithm, we use the  correlation coefficient between
 the estimated and exact channel vector, given by
\begin{equation}
\begin{aligned}
\eta_{k} &= \frac{|\B{h}_k^{\rm H} \hat{\B{h}}_k |}{\left\|\B{h}_k  \right\|_2\left\|\hat{\B{h}}_k  \right\|_2}  \\
   &=  \frac{|\B{h}_k^{\rm H} (\B{h}_k+\B{e}_k) |}{\left\|\B{h}_k  \right\|_2\left\| \B{h}_k + \B{e}_k  \right\|_2} \\
   &=\frac{|\B{h}_k^{\rm H} (\B{h}_k+\B{e}_k) |}{\left\|\B{h}_k  \right\|_2\left\|  \B{e}_k  \right\|_2} \cdot \sqrt{\frac{\left\|\B{h}_k  \right\|_2\left\|  \B{e}_k  \right\|_2}{ \left\| \B{h}_k + \B{e}_k  \right\|_2^2}}  \cdot \sqrt{\frac{\left\|  \B{e}_k  \right\|_2}{ \left\| \B{h}_k \right\|_2}},
   \end{aligned}
\end{equation}
where $\B{e}_k$  denotes the channel estimation error.  As a theoretical performance benchmark, we consider the following expression based on the Fisher information matrix in (\ref{fisher}) and (\ref{fisher2})
\begin{equation}
\begin{aligned}
 \eta_{{\rm CRB},k}  & \approx \frac{\left\|\B{h}_k  \right\|_2}{\sqrt{\left\|\B{h}_k  \right\|_2^2+\left\|\B{e}_k  \right\|_2^2}}  \\
   & \approx \frac{1}{\sqrt{1+  \frac{\sum\limits_{i=1}^{L_k} \left[\tilde{\B{J}}^{-1}\right]_{\sum_{k'=1}^{k-1}L_{k'}+i,\sum_{k'=1}^{k-1}L_{k'}+i }}{\left\|\B{h}_k  \right\|_2^2}}} ,
\end{aligned}
\end{equation}
where we neglect the correlation factor between the estimation error $\B{e}$ and the channel vector $\B{h}$ for the unbiased estimator in the large system limit, i.e.,
\begin{equation}
\lim\limits_{N\rightarrow \infty} \frac{|\B{h}_k^{\rm H} \B{e}_k |}{\left\|\B{e}_k  \right\|_2 \left\|\B{h}_k  \right\|_2} \longrightarrow 0.
\end{equation}
Further, we approximate the squared norm of $\B{e}$ by the Cram\'er Rao Bound due to the law of large numbers.
In the simulation scenario, we use $K=2$ users, $N=32$ antennas, $L_k=3$ multi-path components and a coherence length of $T=1000$. The empirical complementary cumulative distribution of the correlation factor among all users and 100 channel realizations obtained by the subspace-based estimation (without sparsity assumption) and the sparsity-based blind estimation is shown in Fig.~\ref{fig_est}. The $\ell_1$ regularization parameter is fixed at $\lambda=4$. We note that the generated  angles of arrival do not necessarily fall on the DFT grid, which means that the sparsity in the discretized angular domain does not perfectly hold. Nevertheless, taking into account the sparsity of the propagation channel improves the performance significantly and approaches the clairvoyant Cram\'er Rao Bound. Concerning the non-uniqueness of the solution with respect to user permutations as described in (\ref{perm}), we chose for both blind methods the permutation maximizing $\eta_1+\eta_2$  in this simulation.  

Furthermore, we compare the proposed approach with the semi-blind approach presented in \cite{Neumann_2015}, where $T_{\rm T}$ symbols, denoted by $\B{X}_{\rm T}$, out of the $T$ sized block are dedicated for training. Thereby sparsity is not taken into account and the maximum likelihood optimization is formulated as (c.f. (\ref{opt_S}))
\begin{equation}
\begin{aligned}
    &\max\limits_{\B{H}} L(\B{H})-  \left\| \B{H}\B{X}_{\rm T}- \B{Y}_{\rm T} \right\|_2^2 = \\
   &\!\! -\! {\rm tr} \! \left(\! \! \B{Y}^{\rm H}_{\rm D } \! \left( \!\rho \B{H}\B{H}^{\rm H} \!+\! \B{I} \!\right)^{-1} \!\!   \B{Y}_{\rm D} \!\! \right) \!-\!(T-T_{\rm T}) \log\left|\rho \B{H}\B{H}^{\rm H}\! +\! \B{I} \right|  \\
   &\quad\quad\quad\quad\quad\quad\quad\quad\quad\quad\quad\quad\quad\quad\quad\quad- \left\| \B{H}\B{X}_{\rm T}- \B{Y}_{\rm T} \right\|_2^2,
\end{aligned}    
\end{equation}
where $\B{Y}_{\rm T}$ and $\B{Y}_{\rm D}$ represent the received signals corresponding to the known training block (commonly orthogonal sequences) and the unknown data block, respectively. For the simulation scenario, we take $K=2$ orthogonal sequences of length $T_{\rm T}=10$ and solve the optimization problem using the gradient based method. Fig.~\ref{fig_est} shows that the proposed pure blind approach only exploiting the sparsity still outperforms the semi-blind approach using a pilot of length $T_{\rm T}=10$. This confirms the usefulness of the presented method in terms of reducing the training overhead. It is also expected that further performance advantages could be obtained with higher number of users and larger antenna arrays.  

\section{Conclusion}
This work considered the maximum likelihood problem of blind massive MIMO channel estimation.  Based on the sparsity in the angular frequency domain, an $\ell_1$ regularized optimization problem is formulated and solved using a fixed point iteration. The method significantly improves the spectral efficiency by dramatically reducing the overhead caused by pilot sequences and only exploits the sparsity property, and it therefore robust to any type of statistical properties of data and channels. Simulations
demonstrate that this maximum likelihood approach can
achieve a dramatic improvement in performance at low SNR. In fact, it allows blind separation of non-orthogonal channels just by taking advantage of the sparsity assumption.

\section*{Appendix}
First consider the subdifferentials of the $\ell_{1,1}$-norm $\left\| \B{S}\right\|_{1,1}$ with respect to any element $s_{i,j}$ using  Wirtinger's calculus. We have
\begin{equation}
  \frac{\partial |z|}{\partial z}= \frac{\partial \sqrt{z\cdot z^*}}{\partial z}=\frac{z^*}{2\sqrt{z\cdot z^*}}=\frac{1}{2} {\rm e}^{-{\rm j} \angle(z)}.
\end{equation}
Therefore, we get the subdifferentials 
\begin{equation}
 \partial_{s_{i,j}} \left\| \B{S}\right\|_{1,1} \in \left\{
\begin{array}{cc}
	\frac{1}{2} {\rm e}^{-{\rm j} \angle(s_{i,j})} & \textrm{for~} s_{i,j} \neq 0, \\
	\{ \frac{1}{2} {\rm e}^{-{\rm j} \phi}  |\forall \phi \} & \textrm{for~} s_{i,j} = 0.
\end{array}\right.
\end{equation}
Then, the KKT conditions of the optimization problem (\ref{opt_S}) can be written as
\begin{equation}
 -\partial_{s_{i,j}} L(\B{S}) \in \left\{
\begin{array}{cc}
	-\frac{\lambda}{2} {\rm e}^{-{\rm j} \angle(s_{i,j})} & \textrm{for~} s_{i,j} \neq 0, \\
	\{ \frac{\lambda}{2} {\rm e}^{-{\rm j} \phi}  |\forall \phi \} & \textrm{for~} s_{i,j} = 0.
\end{array}\right.
\label{KKT}
\end{equation}
Considering now the following fixed point equation
\begin{equation}
\begin{aligned}
 &\B{S}= \\
 &{\rm exp} ( {\rm j} \angle (\B{S} -\mu \B{\Delta})) \circ \max \left( {\rm abs}( \B{S} -\mu \B{\Delta} ) - \mu \frac{\lambda}{2} \B{1} \cdot \B{1}^{\rm T}, \B{0} \right),
 \end{aligned}
 \label{fixed}
\end{equation}
with $\B{\Delta}^*=-\nabla_{\B{S}}L(\B{S})$ and $\mu>0$, any of its solutions is also a solution of the KKT conditions (\ref{KKT}). Solving (\ref{fixed}) using the fixed point iteration (\ref{fixed_iter}) with $\mu$ small enough converges provided that the gradient $\Delta$ is bounded and yields a local optimum for the optimization problem (\ref{opt_S}).

\section*{Acknowledgment}

This work was supported by the National Science Foundation under
ECCS-1547155, and by the Technische Universit\"at M\"unchen Institute for
Advanced Study, funded by the German Excellence Initiative and the European
Union Seventh Framework Programme under Grant No. 291763, and
by the European Union under the Marie Curie COFUND Program.



%
\bibliographystyle{plain}     
\bibliography{references}{}

\end{document}